\documentclass[conference]{IEEEtran}
\IEEEoverridecommandlockouts

\usepackage{cite}
\usepackage{amsmath,amssymb,amsfonts}
\usepackage{algorithmic}
\usepackage{graphicx}
\usepackage{textcomp}
\usepackage{xcolor}
\usepackage{mathtools}
\usepackage{bm}
\usepackage{comment}
\usepackage{subfig}
\usepackage{wrapfig}
\usepackage{booktabs} 

\usepackage[font=small,labelfont=bf]{caption}

\usepackage{array}
\newcolumntype{P}[1]{>{\centering\arraybackslash}p{#1}}
\newcolumntype{M}[1]{>{\centering\arraybackslash}m{#1}}

\def\BibTeX{{\rm B\kern-.05em{\sc i\kern-.025em b}\kern-.08em
    T\kern-.1667em\lower.7ex\hbox{E}\kern-.125emX}}

\begin{document}

\bstctlcite{MyBSTcontrol}

\title{A Cross-Platform Analysis of Practical Quantum
Error Correction Codes}


\author{
  \IEEEauthorblockN{Yufeng Xin and Bryan Pan}
  \IEEEauthorblockA{RENCI, University of North Carolina at Chapel Hill\\
    Chapel Hill, NC, USA\\
    yxin, panb@renci.org}
}

\maketitle

\begin{abstract}
The theory of quantum error correction was established decades ago. Yet limitations in physical noise levels and available physical-qubit counts continue to hinder the development of scalable quantum computing systems. In this paper, we present analytical estimates of logical error rates for advanced QEC codes across leading hardware platforms and distributed quantum computing systems using a simple but unified framework. The analysis captures dominant contributors to logical error, including effective circuit volume and two-qubit gate overhead. The framework provides a fast and conservative estimate of logical error rates and identification of dominating factors in different hardware platforms, such as circuit volume, routing overhead, inter-QPU operations, or asymmetric noise protection. We show that several qualitative trends observed in larger-scale simulations can be reproduced and interpreted analytically within this framework. We further demonstrate that the framework can be used to find the sweet spot design region of distributed QEC, which is critical for the design of distributed quantum computing systems.
\end{abstract}

\section{Introduction}
\label{sec:introduction}
Efficient quantum error correction (QEC) is key to achieving fault tolerance for large-scale quantum computing (QC) systems. However, due to heterogeneity of leading quantum processing unit (QPU) hardware platforms in terms of capacity, connectivity, and error profiles of physical qubits, analysis of QEC codes has been challenging. While distributed quantum computing (DQC) has gained much research attention ~\cite{chen2026qtest,xin:qdcn:25}, analysis of QEC on distributed platforms has been lacking.

In this paper, we present an analytical framework for estimating logical error rates of advanced QEC codes based on a simple probabilistic model. It captures two dominant contributors to logical errors: code structure and two-qubit gate overhead. The former is characterized by its coding rate and distance, and the latter depends on the physical qubit connectivity of hardware platforms. QPUs without fully connected physical qubit topology need extra SWAP gates, which will increase the number of two-qubit gates and thus the logical error rate. This is vital to the performance of QEC codes in distributed quantum computing systems, where inter-QPU communication links are much noisier. 

QEC differs fundamentally from classical error correction: quantum states are continuous non-cloneable and collapsing upon direct measurement. Furthermore, quantum errors include both bit flip and phase flip components, as well as their combinations and coherent superpositions. Thus, QEC must protect quantum information indirectly by encoding a logical qubit into an entangled state of many physical qubits and extracting error information and conducting correction through syndrome measurements and decoding without disturbing the encoded state via ancilla qubits.~\cite{WoottersZurek1982NoCloning, NielsenChuang2010, Gottesman1997Stabilizer, Roffe_2019}.

The vast QEC code design space makes it difficult to compare the fault tolerance capability across heterogeneous quantum platforms. A recent study evaluates major QEC code classes on leading hardware platforms using full-stack quantum circuit simulation~\cite{Swierkowska2025ECCentric}. While it provides highly detailed and realistic quantification, this approach faces fundamental limitations in expensive computational resources, obscure insights mapping the large code space and the hardware constraints, and difficult exploration of new code designs. 

In contrast, analysis based on our lightweight analytical framework aims to provide a fast estimate of logical error rates and identification of
dominating factors. Through extending the basic binomial qubit error probability model, it abstracts the platform with key circuit-level parameters such as two-qubit gate count, physical error rate, and effective fault threshold. The model is readily extended to distributed QEC after explicitly separating intra-QPU and inter-QPU error sources, allowing for identifying tradeoff between increased connectivity and increased physical error rate on inter-QPU links. Additionally, we consider heterogeneous architectures with multiple classes of operations, further broadening the applicability of our framework. 

The key contribution of the paper is the feasible region analysis of nine advanced QEC codes across three leading QPU hardware platforms and distributed quantum computing systems with a joint analytical framework. In addition to the logical error rate estimates, the framework identifies dominating factors across different hardware platforms, such as circuit volume, routing overhead, inter-QPU operations, or asymmetric noise protection. We show that qualitative trends observed in larger-scale simulations can be reproduced and interpreted analytically with this framework. We specifically demonstrate that the framework can be used to find the sweet spot in the design region of distributed QEC implementation. We emphasize that our analysis includes QEC codes that have not been covered in the referred simulation study.

The remainder of this paper is organized as follows. In Section \ref{sec:problem}, we first discuss advanced QEC codes and their structural differences. We then abstract leading hardware platforms and noise models. In Section \ref{sec:solution}, we present our joint analytical framework with a focus on the distributed QEC. Section \ref{sec:evaluation} applies this framework to compare QEC codes across hardware platforms and an extension with a biased noise model. We conclude the paper in Section~\ref{sec:future}.


\section{QEC Codes and QC Hardware Platforms}
\label{sec:problem}
QEC codes can be concisely represented by $ [[n, k, d]] $ notation, where $ n $ is the number of physical qubits, $ k $ is the number of logical qubits, and $ d $ is the code distance. The code distance is defined as the minimum number of physical qubits in error to cause a logical error. At the code level, a QEC code of distance $ d $ can correct up to $ t = \left\lfloor\frac{d - 1}{2}\right\rfloor $ data-qubit errors, giving the code-level threshold $ \kappa_\text{code} = t + 1. $ We use $ \kappa $ to denote the effective circuit-level minimum number of physical faults required for logical failure. When circuit-level propagation information is unavailable, we use the baseline approximation $ \kappa = \kappa_\text{code}. $ For circuits in which a single physical fault can propagate to several data errors or corrupt multiple inferred syndrome constraints, $ \kappa $ must instead be interpreted as an effective parameter or used together with the structural fault-exposure model introduced in Section~\ref{sec:solution}. The coding rate is defined as $ R = \frac{k}{n} $. 

Stabilizers are operators that define and protect logical subspaces of a QEC code space, in which logical qubits are encoded with redundant physical qubits. A stabilizer code is constructed by a set of mutually commuting Pauli operators whose joint $ +1 $ eigenspace encodes the logical qubits. Each stabilizer acts on a subset of physical qubits. When an error occurs, it anticommutes with one or more stabilizers, causing their measurement outcomes to flip from $ +1 $ to $ -1. $ These measurement results (the error syndrome) consist of information about the location of errors without directly measuring or disturbing the encoded quantum state.  A decoder can then repeatedly measure the syndrome and correct errors without disturbing the logical information.

In general, the preferred code is not the one with the best asymptotic properties on paper, but the one whose stabilizer structure aligns with the connectivity and noise characteristics of the underlying hardware. Distributed QC systems introduce more complex tradeoffs between the increased connectivity and noisier inter-QPU links. In this case, the preferred code is typically the one that can achieve the best error suppression with the fewest inter-QPU operations.

\subsection{Practical QEC Codes}
\label{sec:codes}
Advanced QEC stabilizer codes are generally categorized into four primary families: topological codes, subsystem stabilizer codes, concatenated codes, and quantum low-density parity-check (qLDPC) codes~\cite{Swierkowska2025ECCentric}. We add Floquet codes as a distinct class characterized by time-dependent stabilizer measurements. 

{\bf 1. Topological codes} encode logical information into a lattice and have both high error thresholds and a natural fit for hardware platforms with local connectivity, such as superconducting circuits. Topological codes typically require a large number of stabilizer measurements per cycle, resulting in high circuit volume but strong error suppression in sufficiently low noise regimes. We will analyze three major variants: surface, color, and $ XZZX $ surface topological codes~\cite{BonillaAtaides2021XZZX}.

{\bf 2. Subsystem Stabilizer Codes} extend stabilizer codes by introducing additional degrees of freedom (gauge qubits), allowing stabilizer measurements to be decomposed into lower weight operators. This reduces measurement complexity and can significantly decrease circuit depth. This construction works well on architectures with flexible connectivity, such as trapped-ion systems. We will analyze three representative codes in this family: Bacon-Shor, Heavy-hex, and 2D Gauge Color codes~\cite{Bombin2015GaugeColorCodes}.

{\bf 3. Concatenated Codes} encode logical qubits recursively by nesting one code within another. The concatenated Steane code is a canonical example, constructed by recursively encoding the $ [[7, 1, 3]]$ Steane code. While concatenation obviously improves error suppression, it leads to exponential growth in qubit count and reduced coding rate. Concatenated codes trade locality for reduced circuit volume, which can make them advantageous in noisy or low connectivity environments. We will analyze the Concatenated Steane code in this paper~\cite{Steane1996Multiple}.

{\bf 4. qLDPC Codes} (Quantum low-density parity-check codes) are a class of stabilizer codes in which each stabilizer generator acts on only a small number of qubits, and each qubit participates in only a small number of stabilizers. Analogous to classical LDPC codes, it enables efficient decoding algorithms. Unlike topological codes, qLDPC codes are not constrained by locality, so they can achieve constant coding rates. However, this advantage does require non-local interactions, so implementing qLDPC codes typically requires long-range connectivity or additional routing overhead that can significantly increase circuit volume on hardware with limited connectivity. We will analyze the Bivariate Bicycle code, the latest qLDPC code with constant encoding rates and relatively high error thresholds~\cite{Bravyi2024BBMemory}.

{\bf 5. Floquet Codes} do not have fixed stabilizer measurements. Instead, they vary periodically in time, measuring sequences of low-weight Pauli operators according to a predefined schedule, from which the effective stabilizer structure forms dynamically. This approach uses only weight-two measurements, significantly reducing circuit depth and two-qubit gate count compared to conventional stabilizer codes. Thus, Floquet codes are able to achieve beneficial tradeoffs between circuit volume and error suppression, particularly in regimes where two-qubit gate errors dominate. In this paper, we specifically analyze the EM3 Floquet variant associated with the Bolza surface construction~\cite{Higgott_2024}.

\subsection{Leading QC Hardware Platforms} 
\label{sec:hardwarenoise}
The logical error is determined by the number of physical operations required to implement a QEC cycle, which is primarily affected by the physical qubit connectivity, in addition to the different noise characteristics. On a fully connected architecture, any two-qubit interaction required by the code can be executed directly, so the physical two-qubit gate count is close to the ideal circuit count. On sparse topologies, non-local interactions must be routed through SWAP networks, increasing the total number of two-qubit gates and therefore increasing possible faults. In this section, we describe the three main platforms with respect to the count of possible physical fault locations $ N_\text{loc} $ and their practical approximate per-location physical error rate $ p_{\text{loc}}. $ 

{\bf 1. Trapped-Ion} systems are characterized by near all-to-all connectivity within a single QPU. As a result, multi-qubit circuits can typically be executed without the need for SWAP-based routing. This high connectivity allows logical circuits to be implemented with minimal overhead, so the physical two-qubit gate count closely matches the logical requirement. Therefore, we approximate $N_\text{loc} \approx N_{2Q} $, where $N_{2Q}$ denotes the number of logical two-qubit gates. Furthermore, trapped-ion systems also achieve very high gate fidelities with error rates often reaching the $ 10^{-4} $ level or below~\cite{Gaebler2016IonGates}. Single-qubit gate and state preparation and measurement (SPAM) errors are typically an order of magnitude lower, and therefore the contribution to logical failure is negligible.

{\bf 2. Superconducting} QPUs face the connectivity constraint that two-qubit gates can only be performed between adjacent qubits, and non-local interactions must be implemented through sequences of SWAP operations, so superconducting qubits incur more connectivity overhead~\cite{Kjaergaard_2020}. To model this, we can distinguish between the logical two-qubit gate count required by a QEC code and the physical two-qubit gate count after routing. Therefore, the total number of fault locations per QEC cycle can be approximated as  $ N_\text{loc} \approx N_{2Q} + N_\text{SWAP} + N_{\text{meas}}, $ where $ N_\text{SWAP} $ denotes the additional routing overhead induced by limited connectivity and $ N_{\text{meas}} $ denotes the number of measurement error locations.

{\bf 3. Neutral-Atom} QPUs can support flexible connectivity through qubit rearrangement and long-range Rydberg-mediated interactions. However, this flexibility often requires additional atom movement, scheduling constraints, or extra entangling operations, all of which can increase the total number of physical fault locations relative to the logical circuit description~\cite{Wang2024Atomique}. Thus, neutral-atom platforms typically incur additional movement and scheduling overhead relative to the idealized logical circuit, although the magnitude of this overhead is highly architecture and compiler dependent. As a result, the effective number of fault locations per QEC cycle may be substantially larger than the logical two-qubit gate count
$ N_\text{loc} \gtrsim N_{2Q}. $ Furthermore, neutral-atom systems currently tend to have noisier gates and more error sources than trapped-ion platforms, so the approximation of $ p_\text{loc} $ with the two-qubit gate fault probability $ p_{2Q} $ is often less accurate than in the trapped-ion case. 

\subsection{Realistic Noise Models}
We consider both the standard symmetric depolarizing model as well as more realistic biased noise models observed in leading quantum hardware.

{\bf 1. Symmetric Depolarizing Noise} is a widely used model due to its simplicity. It assumes that each fault location fails independently with probability $ p_\text{loc}. $ When a fault occurs, it results in a Pauli error drawn uniformly from $ \{X, Y, Z\}. $ Thus, bit flip and phase flip components occur symmetrically, although they are not strictly independent because a $ Y $ error contains both $ X $ and $ Z $ components. Under this assumption, no error type is favored over another. This model is widely used due to its simplicity, but it does not accurately reflect the behavior of most physical quantum systems, where different error channels often occur with different probabilities~\cite{Roffe_2019}. 

{\bf 2. Biased Noise.}
In particular, many qubit platforms typically exhibit strongly biased noise where phase errors ($ Z $ errors) occur far more frequently than bit-flip errors ($ X $ errors)~\cite{BonillaAtaides2021XZZX}. For simplicity, we neglect explicit $ Y $-type error correlations. This approximation is sufficient for comparing how different effective $ X $ and $ Z $ distances respond to increasing bias. We represent this kind of noise using a new parameter $ \eta = \frac{p_Z}{p_X} $ where $ p_Z $ and $ p_X $ denote the probabilities of phase and bit-flip errors respectively. Under biased noise, the total physical error probability is simply $ p_\text{loc} = p_X + p_Z. $ We can also represent $ p_X $ and $ p_Z $ as functions of $ \eta $ by
$$ p_X(\eta) = \frac{p_\text{loc}}{1 + \eta}, \quad p_Z(\eta) = \frac{\eta p_\text{loc}}{1 + \eta}. $$ 
It is well motivated to study moderate bias values such as $ \eta \sim 5-10, $ as well as larger bias regimes in engineered architectures~\cite{Etxezarreta2025CircuitBiasedNoise}.

{\bf 3. Heterogeneous Physical Error Rates.}
The previous models assume that every fault location has the same error probability. Real hardware, however, typically has non-uniform error rates across qubits, couplers, and measurement channels. To model this, we let each physical location $ i $ fail independently with probability $ p_i. $ The total number of faults is then a Poisson-binomial random variable rather than just a standard binomial random variable:
$$ X = \sum_{i = 1}^{N_\text{loc}} X_i, \quad X_i \sim \mathrm{Bernoulli}(p_i). $$

This distribution is more general, but its leading order behavior is still controlled by the total expected number of faults,
Thus, if two devices have the same average physical error rate and comparable variance, their logical error rates may be similar even if the individual qubits have different fidelities. This explains why non-uniform physical error distributions typically have a weaker effect than large changes in circuit volume or average error scale.

\section{Analytical Model for Logical Error Rate Estimation}
\label{sec:solution}
In our analytical model, factors contributing to the logical error rate are abstracted through two variables: the effective number of fault locations per correction cycle $ N_\text{loc} $ and the effective physical error probability per location, $ p_\text{loc}. $ For a code $ [[n, k, d]], $ increasing $ d $ improves error suppression by increasing $ \kappa, $ while hardware and implementation effects worsen performance by increasing either $ N_\text{loc} $ or $ p_\text{loc}. $

We can decompose the effective circuit volume as
$$ N_\text{loc} = N_\text{native} + N_\text{SWAP} + N_\text{meas} + N_\text{idle} + N_\text{inter}, $$
where $ N_\text{native} = N_{2Q} $ is the number of native two-qubit operations required by the QEC circuit, $ N_\text{SWAP} $ is the additional routing overhead induced by limited qubit connectivity, $ N_\text{meas} $ is the number of measurement-related fault locations, $ N_\text{idle} $ accounts for memory or idle errors during syndrome extraction, and $ N_\text{inter} $ counts inter-QPU operations in distributed architectures. In the single-QPU trapped-ion case, $ N_\text{SWAP} $ and $ N_\text{inter} $ are negligible, so $ N_{\text{loc}} \approx N_\text{native} = N_{2Q}. $ In connectivity-limited or distributed systems, however, these additional terms can significantly increase the logical error rate.

\subsection{Base Model}
\label{subsec:basemodel}
Assuming the symmetric depolarizing noise model with independent fault location (a two-qubit gate, measurement, etc.) failure with probability $ p_\text{loc} $~\cite{AharonovBenOr2008, WallmanEmerson2016, NielsenChuang2010,  Roffe_2019}, the number of faults $ X $ in a QEC cycle follows a binomial distribution $ X \sim \mathrm{Binomial}(N_\text{loc}, p_\text{loc}). $  We further assume that $ N_{2Q} $ is a sufficient leading-order proxy for circuit size and error exposure. Then the binomial tail probability is a first-order estimate for the logical error probability:
\begin{equation} \label{eqn:binomsingleqpu}
p_L \leq \mathbb{P}[X \geq \kappa] = \sum_{j = \kappa}^{N_\text{loc}} {N_\text{loc} \choose j}(p_\text{loc})^j(1 - p_\text{loc})^{N_\text{loc} - j}.
\end{equation}
This base model essentially isolates the dominant contributors to logical failure, namely circuit volume and two-qubit gate fidelity under the assumption that two-qubit gates are the primary error source. Other significant sources of errors are ignored: stochastic Pauli error channels can be suppressed using dynamical decoupling, coherent errors are randomized away under twirling, SPAM errors are mitigable assuming measurement errors are not strongly or densely correlated, and ideally relaxation effects are small enough to be neglected.

\subsection{Distributed QPUs}
To extend our analysis to distributed codes, we start by assuming multiple identical QPUs and partition fault locations per cycle into intra-QPU and inter-QPU classes~\cite{PhysRevX.4.041041, Ramette2024FaultTolerant, sutcliffe2025distributedquantumerrorcorrection}. We let $ N_\text{intra} $ be the number of intra-QPU $ 2Q $ gates and $ N_\text{inter} $ be the number of inter-QPU $ 2Q $ gates. Similarly, $ p_\text{intra} $ and $ p_\text{inter} $ are the intra-QPU and inter-QPU $ 2Q $ gate errors respectively. Like in the single QPU case, we approximate $ p_L $ by the probability that at least $ \kappa $ faults occur in one cycle. The number of inter-QPU faults and intra-QPU faults can also be represented by binomial distributions
$$ X \sim \mathrm{Binomial}(N_\text{inter}, p_\text{inter}), \quad Y \sim \mathrm{Binomial}(N_\text{intra}, p_\text{intra}) $$
and the total faults is simply $ Z = X + Y. $ Our upper bound for logical error assuming failure for at least $ \kappa $ faults is then
\begin{equation} \label{eqn:binomdistributed}
p_L \leq \sum_{m = \kappa}^{N_\text{inter} + N_\text{intra}}\sum_{j = \max(0, m - N_\text{intra})}^{\min(m, N_\text{inter})}\mathbb{P}[X = j]\mathbb{P}[Y = m - j].
\end{equation}

In practice, distributed quantum systems may consist of heterogeneous hardware modules with different physical error characteristics. For example, a system may combine trapped-ion modules with superconducting processors, or use photonic interconnects whose error rates differ significantly from local two-qubit gates. 
To account for this, suppose a QEC cycle contains $ C $ distinct classes of fault locations. What each class corresponds to is arbitrary, but some examples are two-qubit gates within QPU $ a, $ two-qubit gates within QPU $ b, $ or interconnect operations between QPU $ a $ and $ b. $  Let $ N_c $ denote the number of fault locations of class $ c $ per cycle, and let $ p_c $ denote the corresponding physical error probability. Assuming all classes are independent, the number of faults contributed by class $ c $ is then just
$ X_c \sim \mathrm{Binomial}(N_c, p_c). $
The total number of faults in one cycle is
$ Z = \sum_{c = 1}^C X_c. $
We then need to calculate $ \mathbb{P}[Z \geq \kappa] $ as the upper bound for $ p_L. $ The probability generating function (PGF) for $ Z $ is 
$$ G(s) = \prod_{c = 1}^{C} ((1 - p_c) + p_c s)^{N_c}. $$
This is equivalent to a grouped Poisson-binomial model, where the total number of faults is the sum of independent Bernoulli or binomial contributions with different error probabilities~\cite{Hong2013PoissonBinomial}. The probability of observing exactly $ m $ faults is the coefficient of $ s^m $ in $ G(s): $
$$ \mathbb{P}[Z = m] = [s^m] G(s), $$
and therefore
\begin{equation} \label{eqn:binomdistributedheterogeneous}
p_L \leq \mathbb{P}[Z \geq \kappa] = \sum_{m = \kappa}^{N_{\text{tot}}} [s^m] G(s), 
N_{\text{tot}} = \sum_{c = 1}^{C} N_c.
\end{equation}
It is useful to note that
$$ \mathbb{P}[\mathrm{Bin}(N_{\text{tot}}, p_{\min}) \geq \kappa] \leq p_L \leq \mathbb{P}[\mathrm{Bin}(N_{\text{tot}}, p_{\max}) \geq \kappa]. $$
where 
$$ p_{\min} = \min_{c} p_c, \quad p_{\max} = \max_{c} p_c. $$
Thus, a heterogeneous system behaves no better than a uniform device with error rate $ p_{\min} $ and no worse than one with error rate $ p_{\max}. $  Without having to calculate our expression for $ p_L $ entirely, we can instead compute
$$ \mathbb{E}[Z] = \sum_{c = 1}^C N_cp_c. $$
If $ \mathbb{E}[Z] \ll \kappa, $ error suppression is possible, but if $ \mathbb{E}[Z] \gtrsim \kappa, $ logical failure becomes likely regardless of code structure. \\

\subsection{Code-dependent structural fault exposure.} \label{subsec:struct}
The physical location count $ N_{\text{loc}}$ measures circuit volume, but it does not by itself describe how strongly a fault at one location can affect the inferred syndrome. This distinction is especially relevant for subsystem codes, in which stabilizers are reconstructed from several gauge measurements, and for Floquet codes, in which stabilizer information is accumulated over a time-dependent measurement schedule. In these constructions, one physical fault may corrupt multiple inferred stabilizer values or multiple temporal syndrome constraints.

To incorporate this effect without requiring a complete full-stack simulation, we partition the circuit into $ C $ classes of fault locations. Class $ c $ contains $ N_c $ locations with physical error probability $ p_c, $ and we associate it with an integer structural propagation weight $ w_c \geq 1. $ The weight $ w_c $ represents a conservative estimate of the number of effective harmful fault units that may be produced by one physical fault in that class. Let
 $X_c$ denote the number of physical faults in class $ c. $ We define the weighted structural fault load as
$$ S = \sum_{c = 1}^{C}w_cX_c. $$
If the weights are selected conservatively so that $ S<\kappa $ implies that logical failure is impossible, then the logical error probability satisfies
$$ p_L^{\text{circ}} \leq \mathbb{P}[S \geq \kappa]. $$
For integer weights, the PGF of $ S $ is
$$ G_w(s) = \prod_{c = 1}^{C} \left[(1 - p_c) + p_cs^{w_c}\right]^{N_c}, $$
and the structural bound can be evaluated from
$$ \mathbb{P}[S \geq \kappa] = \sum_{m = \kappa}^{S_{\max}}[s^m]G_w(s), \quad S_{\max} = \sum_{c = 1}^{C}w_cN_c. $$

A simpler scalar approximation is obtained by defining the structural exposure multiplier 
$$ \chi_{\text{str}} = \frac{\sum_{c = 1}^{C}w_cN_c}{\sum_{c = 1}^{C}N_c} $$
and the effective location count
 $N_{\text{loc}}^{\text{eff}} = \left\lceil\chi_{\text{str}}N_{\text{loc}}\right\rceil. $
Under a homogeneous physical error rate, replacing $ N_{\text{loc}} $ by $ N_{\text{loc}}^{\text{eff}} $ in the binomial model matches the mean weighted fault exposure. This scalar replacement is a first-order approximation of correlated propagation as several syndrome effects may originate from the same physical fault.

For subsystem and Floquet circuits, the weights may be estimated from check weight, ancilla fan-out, the number of inferred stabilizers affected by one fault, the number of affected time layers, or small-scale circuit fault-injection experiments. Thus, $ \chi_{\text{str}} $ provides a code-dependent hyperparameter that can be varied or calibrated before a complete circuit-and-decoder simulation is available.


\section{Analysis}
\label{sec:evaluation}
We use the ECCentric benchmark instances in Table~\ref{tab:qec_codes} to evaluate QEC codes in different QPU platforms with symmetric and biased noise, and a distributed QPU setting.
\begin{table}[h]
\centering
\begin{tabular}{lccc}
\hline
Code & Distance $ d $ & $ N_{2Q} $ & $ \kappa = t + 1 $ \\
\hline
Surface & 11 & 1320 & 6 \\
Bacon-Shor & 11 & 1320 & 6 \\
Color & 11 & 1440 & 6 \\
Heavy-hex & 11 & 1920 & 6 \\
Bivariate Bicycle (Gross) & 12 & 2592 & 6 \\
Concatenated Steane & 9 & 961 & 5 \\
$ XZZX $ Surface & 11 & 1320 & 6 \\
Gauge Color (2D) & $ \approx 11 $ & 1092 & 6 \\
Floquet (EM3, Bolza $ m = 10 $) & 14 & 2400 & 7 \\
\hline
\end{tabular}
\caption{QEC Code Parameters}
\label{tab:qec_codes}
\vspace{-0.1in}
\end{table}

\subsection{Single QPU}
Unless otherwise stated, the single-QPU estimates use $ N_\text{loc} = N_{2Q} $ to isolate the effect of the physical two-qubit error rate. We obtained logical error rate estimates on four representative hardware platforms in Table~\ref{tab:singleqpu}: trapped-ion hardware, using Quantinuum Apollo with $ p_{2Q} = 1.4 \times 10^{-4}; $ superconducting hardware, using Willow with $ p_{2Q} = 2.8 \times 10^{-3}; $ and neutral-atom hardware, using Flamingo with $ p_{2Q} = 2.0 \times 10^{-3} $ and Infleqtion with $ p_{2Q} = 6.5 \times 10^{-3}. $ 
\begin{table*}[h]
\centering
\begin{tabular}{lcccc}
\hline
Code & Q. Apollo $ p_L $ per cycle & Willow $ p_L $ per cycle & Flamingo $ p_L $ per cycle & Infleqtion $ p_L $ per cycle \\
\hline
Gauge Color & $ 1.54 \times 10^{-8} $ & $ 8.96 \times 10^{-2} $ & $ 2.40 \times 10^{-2} $ & $ 7.12 \times 10^{-1} $ \\
Surface & $ 4.67 \times 10^{-8} $ & $ 1.69 \times 10^{-1} $ & $ 5.19 \times 10^{-2} $ & $ 8.57 \times 10^{-1} $ \\
$ XZZX $ & $ 4.67 \times 10^{-8} $ & $ 1.69 \times 10^{-1} $ & $ 5.19 \times 10^{-2} $ & $ 8.57 \times 10^{-1} $ \\
Bacon-Shor & $ 4.67 \times 10^{-8} $ & $ 1.69 \times 10^{-1} $ & $ 5.19 \times 10^{-2} $ & $ 8.57 \times 10^{-1} $ \\
Floquet (EM3, Bolza $ m = 10 $) & $ 7.10 \times 10^{-8} $ & $ 5.08 \times 10^{-1} $ & $ 2.09 \times 10^{-1} $ & $ 9.95 \times 10^{-1} $ \\
Color & $ 7.77 \times 10^{-8} $ & $ 2.20 \times 10^{-1} $ & $ 7.21 \times 10^{-2} $ & $ 9.05 \times 10^{-1} $ \\
Steane & $ 3.25 \times 10^{-7} $ & $ 1.35 \times 10^{-1} $ & $ 4.57 \times 10^{-2} $ & $ 7.47 \times 10^{-1} $ \\
Heavy-hex & $ 4.13 \times 10^{-7} $ & $ 4.50 \times 10^{-1} $ & $ 1.90 \times 10^{-1} $ & $ 9.85 \times 10^{-1} $ \\
Bivariate Bicycle (Gross) & $ 2.31 \times 10^{-6} $ & $ 7.31 \times 10^{-1} $ & $ 4.16 \times 10^{-1} $ & $ 9.99 \times 10^{-1} $ \\
\hline
\end{tabular}
\caption{Logical error estimates per cycle for different codes for trapped-ion, superconducting, and neutral-atom platforms.}
\label{tab:singleqpu}
\vspace{-0.1in}
\end{table*}

\begin{figure}
\centering
\includegraphics[width=0.95\columnwidth]{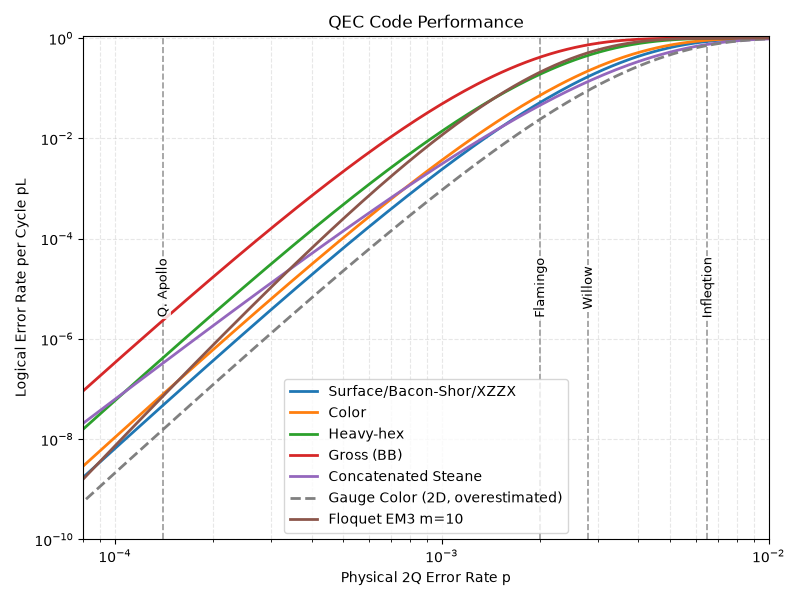}
\caption{Logical error estimates per round vs. two-qubit error rate}
\label{fig:allcodes}
\vspace{-0.1in}
\end{figure}

The results are consistent with the expected leading-order relationship between circuit volume and logical error suppression. Across the three hardware types, codes with smaller $ N_{2Q} $ generally achieve lower logical error estimates because most benchmarked codes have similar $ \kappa. $ Gauge Color gives the lowest estimate due to its smaller circuit volume, while Heavy-hex, Bivariate Bicycle, and the higher-volume Floquet instance have larger estimates because their increased circuit volume outweighs their comparable or larger code distances. Surface, $ XZZX, $ and Bacon-Shor have identical estimates in the base model because they share the same $ N_{2Q} $ and $ \kappa. $ More generally, any codes with identical $ N_{\text{loc}}, p_{\text{loc}}, $ and $ \kappa $ collapse to the same prediction under the unweighted binomial abstraction. Section~\ref{subsec:struct} can distinguish such codes by assigning different propagation weights or different values of $ \chi_{\text{str}}. $ Fig.~\ref{fig:allcodes} further illustrates that the relative ordering of codes remains largely consistent in the low-error regime.

To compare our analytical model with ECCentric, we evaluate the binomial model at $ p_{\text{loc}} = 0.004, $ matching the physical error rate used in its experiments~\cite{Swierkowska2025ECCentric}. The analytical model predicts logical error rates in the range $ \sim 0.3-0.9, $ with an average of approximately $ 0.57 $ across the codes considered. In contrast, ECCentric reports an average logical error rate of approximately $ 0.22 $ for Apollo hardware. At the matched physical error rate, our analytical estimate is therefore more pessimistic for these benchmark instances.

In a complete syndrome-extraction circuit, however, only a subset of these fault combinations is actually malignant. Some faults occur on ancillas or measurements without producing harmful data errors, some physical errors cancel or belong to degenerate error classes, and some apparently large fault patterns are still resolved successfully by the decoder. Other fault sets may contain at least $ \kappa $ faults without forming a logical operator. Consequently, for a fixed set of modeled fault locations and provided that $ \kappa $ is the effective circuit-level minimum number of faults required for failure, the circuit-level logical error probability satisfies
$$ p_L^{\text{circ}} \leq \mathbb{P}[X\geq\kappa]. $$
In this restricted sense, the binomial tail is a decoder-agnostic structural upper bound: it deliberately treats all fault combinations above the threshold as harmful and does not credit the code for error degeneracy, benign fault placement, cancellation, or successful decoding.

This upper-bound interpretation applies only to the circuit abstraction represented by $ N_{\text{loc}} $ and $ \kappa. $ If gauge-check inference, hook-error propagation, or a time-dependent measurement schedule allows fewer than the nominal $ t + 1 $ physical faults to produce a logical failure, then $ \kappa $ must be replaced by an effective circuit-level value or the structural weighting extension introduced in Section~\ref{subsec:struct} must be used.

There are therefore two competing approximations in the base model. Treating every event with at least $ \kappa $ faults as a failure is pessimistic. In contrast, setting $ N_{\text{loc}} = N_{2Q} $ is optimistic because it omits routing, measurement, reset, idle-time, movement, and other implementation-dependent fault locations. For highly connected trapped-ion systems, where these additional overheads are relatively small, the pessimistic threshold assumption may dominate. For connectivity-limited superconducting and neutral-atom platforms, the optimistic undercounting of $ N_{\text{loc}} $ may instead become the larger effect. 

The discrepancy with ECCentric is therefore expected. ECCentric emulates mapping, routing, gate decomposition, syndrome structure, fault propagation, and decoder behavior, whereas the analytical model intentionally compresses these effects into a small number of effective parameters. The variation across ECCentric topologies can still be interpreted within our framework as a topology-dependent scaling of $ N_{\text{loc}}. $ Thus, while the analytical model does not reproduce all circuit and decoder-level effects, it captures the dominant scaling with circuit volume and physical error rate. It should therefore be interpreted as a leading-order predictor of logical error trends

\subsection{Biased Noise}
The response of a quantum error-correcting code to biased noise depends on its effective distances against $ X $ and $ Z $-type logical operators. For many symmetric stabilizer codes, the effective distances against $ X $ and $ Z $ errors are comparable or identical, so $ d_X = d_Z = d $. For the symmetric standard surface code, color codes, and concatenated Steane codes, biased noise does not change the number of faults required for logical failure. Certain codes can exploit biased noise by effectively increasing their protection against the dominant error type~\cite{Tuckett2019BiasedNoise, BonillaAtaides2021XZZX}. One example is the $ XZZX $ surface code, whose stabilizer structure rotates the conventional surface code lattice so that dominant phase errors align with longer logical operators. When the effective distance against phase faults
$ d_Z^{\text{eff}} > d, $
the corresponding fault threshold increases
$$ \kappa_Z^{\text{eff}} = \left\lfloor\frac{d_Z^{\text{eff}} - 1}{2} \right\rfloor + 1. $$
This leads to reduced binomial tail associated with $ Z $ faults and can therefore significantly reduce the logical error rate when $ \eta $ is large. Subsystem constructions such as Bacon-Shor codes can also benefit from biased noise when their lattice dimensions are chosen asymmetrically so that $ d_Z > d_X $~\cite{AliferisPreskill2008BaconShorBiased}. For comparison, we use the following parameters in Table~\ref{tab:biasedparams}.

\begin{table}[h]
\centering
\begin{tabular}{lcccc}
\hline
Code & $ N_{\text{loc}} $ & $ d_X $ & $ d_Z $ & $ (\kappa_X, \kappa_Z) $ \\
\hline
Surface & 1320 & 11 & 11 & (6, 6) \\
Gauge Color & 1092 & 11 & 11 & (6, 6) \\
Bacon-Shor (asymmetric) & 1320 & 7 & 15 & (4, 8) \\
$ XZZX $ Surface & 1320 & 11 & 19 (effective) & (6, 10) \\
\hline
\end{tabular}
\caption{Parameters in the biased-noise logical error model. }
\label{tab:biasedparams}
\vspace{-0.1in}
\end{table}

\begin{figure}
\centering
\includegraphics[width=0.95\columnwidth]{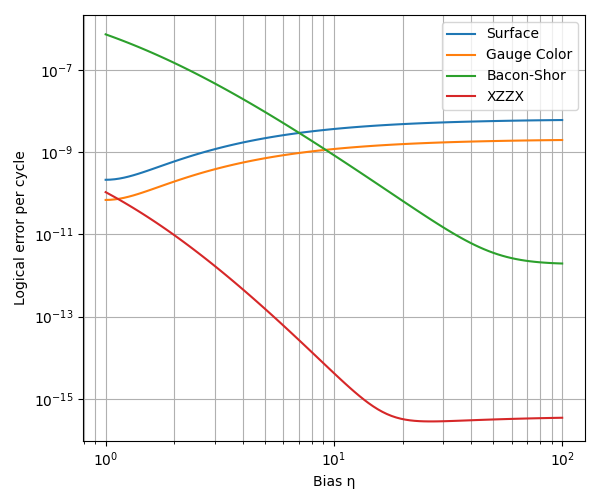}
\caption{$ p_L $ vs bias $ \eta $ with $ p_{\text{loc}} = 10^{-4}. $}
\label{fig:etaplot}
\vspace{-0.1in}
\end{figure}

Figure~\ref{fig:etaplot} presents four codes (Surface, Gauge Color, tailored Bacon-Shor, and $ XZZX $) against $ \eta. $ We choose these four codes as representative examples of the main design classes relevant under biased noise: conventional symmetric topological codes (Surface), conventional symmetric non-surface stabilizer codes (Gauge Color), explicitly asymmetric subsystem codes (Bacon–Shor), and bias-tailored topological codes $ (XZZX). $ It is clear that codes tailored to more $ Z $ errors by having a higher $ Z $ distance vastly outperform traditional codes at high levels of $ \eta. $ For systems that have $ \eta \sim 10 $ or higher, these results suggest that biased codes are extremely attractive. 

\subsection{Distributed QEC with Multiple QPUs}
\label{sec:dist}
In contrast to the fixed-instance single-QPU comparison, the distributed studies below allow the code distance to increase subject to a simplified per-QPU capacity constraint. Any benefit from distribution should therefore be interpreted as capacity-enabled distance growth rather than an intrinsic advantage of distributing a fixed-distance code. We assume that the distributed QC platform consists of $ q $ identical QPUs, each hosting a subset of the code patch, with stabilizer checks that cross QPU boundaries, resulting in inter-QPU operations. According to Equation~\ref{eqn:binomdistributed}, the index $ j $ counts the number of inter-QPU faults among the $ \kappa $ total faults. The purely intra-QPU contribution is the $ j = 0 $ term
$$ T_0 = {N_\text{intra} \choose \kappa} p_\text{intra}^\kappa, $$
while the first correction involving one inter-QPU fault is
$$ T_1 = {N_\text{inter} \choose 1}{N_\text{intra} \choose \kappa - 1}p_\text{inter}p_\text{intra}^{\kappa - 1}. $$
The ratio of these two terms is
$$ \frac{T_1}{T_0} = N_\text{inter} \frac{\kappa}{N_\text{intra} - \kappa + 1} \frac{p_\text{inter}}{p_\text{intra}},
$$
showing that even a small number of inter-QPU operations can substantially increase the logical failure probability when $ p_\text{inter} \gg p_\text{intra}. $ In particular, the first inter-QPU correction remains small only if
$ \frac{T_1}{T_0} \ll 1. $
Typically, $ N_\text{intra} \gg \kappa, $ so this condition simplifies to
$$ N_\text{inter} \ll \frac{N_\text{intra}}{\kappa}\frac{p_\text{intra}}{p_\text{inter}}. $$
With current quantum technology, inter-QPU two-qubit gates are significantly noisier than intra-QPU gates, leading to the realistic assumption of $ p_\text{inter} = p $ and $ p_\text{intra} = p/10 $~\cite{Swierkowska2025ECCentric}.  
Thus, our original leading-order expression can be approximated as
\begin{equation}
\label{eqn:distributed}
p_L \approx \left(\frac{p}{10}\right)^\kappa \sum_{j = 0}^\kappa {N_\text{inter} \choose j} {N_\text{intra} \choose \kappa - j} 10^j.
\end{equation}
The factor $ 10^j $ shows that terms containing more inter-QPU faults are weighted exponentially heavier than terms containing only intra-QPU faults. These contributions are still scaled by the combinatorial factors, so whether the logical error is dominated by small or large $ j $ depends on the balance between the gate counts and the error rate ratio. In the more realistic regime where inter-QPU gates are much fewer than intra-QPU gates, the dominant corrections typically come from small $ j. $

The primary motivation for distribution is not improved error suppression, but scalability. Monolithic QPUs are limited by current hardware technology and therefore cap the maximum achievable device size in terms of available physical qubits. Distributed architectures allow quantum systems to surpass these monolithic limits by interconnecting multiple smaller QPUs.  Distributed execution can be advantageous for architectures or code families that are naturally modular, such as concatenated codes or certain Floquet constructions, where most check operations are local and inter-QPU operations appear only at higher layers or at sparse interfaces~\cite{PhysRevX.4.041041, Ramette2024FaultTolerant, sutcliffe2025distributedquantumerrorcorrection}. 

For heterogeneous distributed hardware, the same conclusion holds: the dominant corrections come from the noisiest fault classes, not necessarily the most numerous ones.

We next use the analytical model for an illustrative scaling study on two distributed QEC codes when each individual QPU has a limited qubit capacity of approximately $ 200 $ physical qubits.

\subsubsection{Distributed Surface-Code Evaluation}
We assume a surface-code-like scaling model $ n_\text{phys}(d) \approx 2d^2$~\cite{Fowler_2012, Ramette2024FaultTolerant}
so that the largest achievable odd distance on $ q $ QPUs is
$$ d_{\max}(q) = \max\{d \text{ odd}: 2d^2 \leq 200q\}. $$
We also use the parameters from the surface code instance in Table~\ref{tab:qec_codes}. Since the $ d = 11 $ surface code circuit has $ N_{2Q} = 1320, $ we approximate $ N_{2Q}(d) \approx 11d^2. $
To model the cost of distribution, we assume that the number of inter-QPU operations grows linearly with QPU count and with the boundary length of the code patch:
$$ N_\text{inter}(q, d) = \alpha d(q - 1) $$
where $ \alpha $ is a constant that represents the number of inter-QPU operations per boundary unit per QEC cycle. In this study, we use $ \alpha = 4 $ as a representative boundary overhead parameter. The remaining two-qubit operations are treated as local intra-QPU operations, so the logical error probability is then estimated as
$$ p_L = \mathbb{P}[X_\text{intra} + X_\text{inter} \geq \kappa]
$$

\begin{figure}[t]
\centering
\includegraphics[width=0.9\columnwidth]{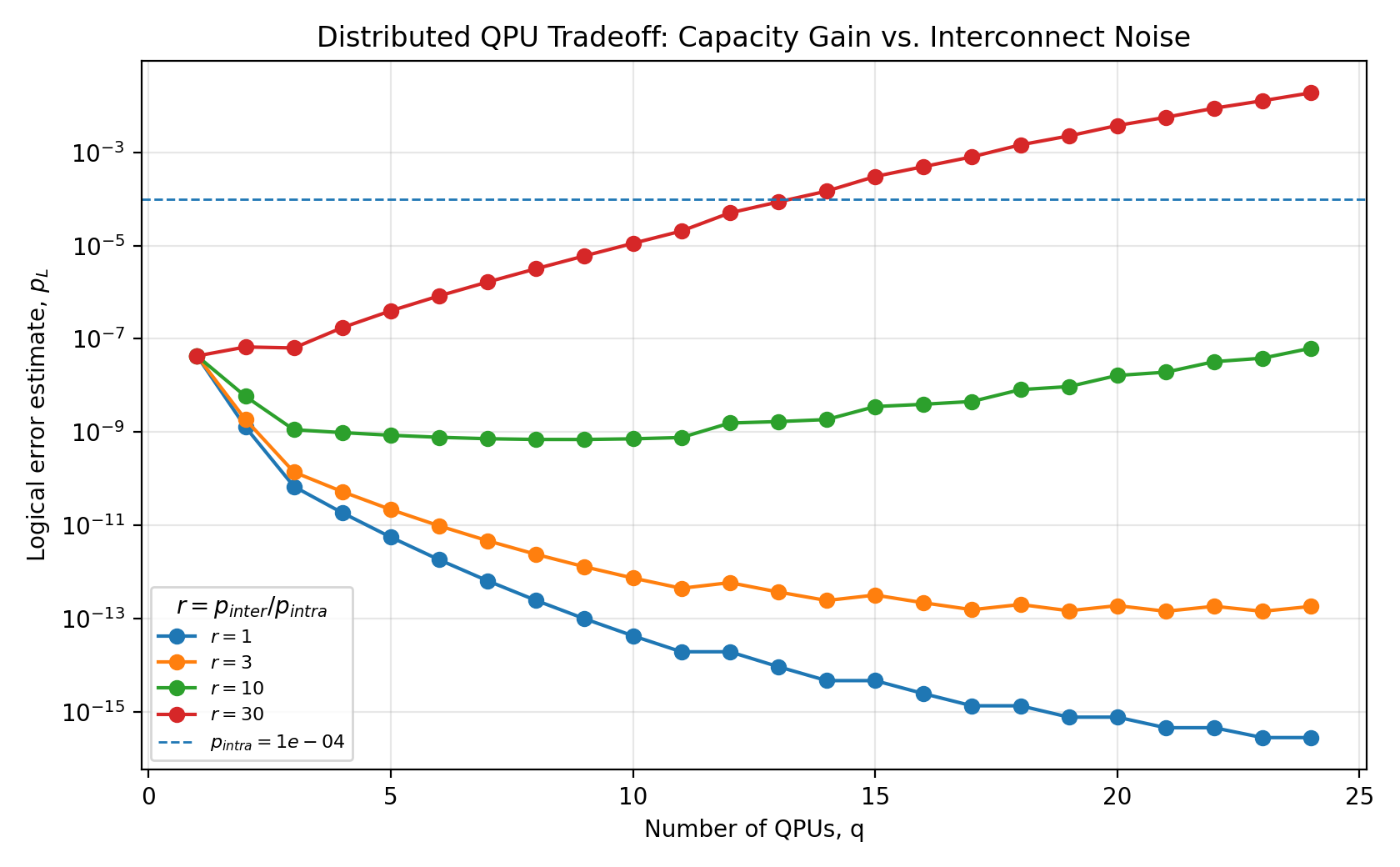}
\caption{Distributed Surface Code logical error estimate: 200 physical qubits per QPU, $ n_\text{phys} \approx 2d^2, N_{2Q} \approx 11d^2, $ and $ N_\text{inter} = 4d(q - 1). $}
\label{fig:distributed_qpu_sweep}
\vspace{-0.1in}
\end{figure}

Figure~\ref{fig:distributed_qpu_sweep} shows the resulting logical error estimates as a function of the number of QPUs for different interconnect noise ratios defined as
$ r = p_\text{inter}/p_\text{intra} $
When inter-QPU links are almost as reliable as local gates, increasing the number of QPUs continues to reduce logical error because the larger achievable code distance dominates the added interconnect overhead. However, as inter-QPU links become noisier, the optimal QPU count decreases.

For the realistic case where $p_\text{intra} = 10^{-4}, \quad r = 10, $ the logical error rate initially decreases as the larger code distance increases the number of correctable faults. However, after approximately $ q = 9, $ the benefit from increasing $d$ is outweighed by the growing inter-QPU overhead. The model predicts a sweet spot around $q = 9, \quad d = 29, \quad \kappa = 15, \quad N_\text{intra} = 8323 $ and $ N_\text{inter} = 928. $ So the expected number of physical faults per cycle is
$ \mu = N_\text{intra}p_\text{intra} + N_\text{inter}p_\text{inter} = 1.76. $
The predicted logical error rate is $ p_L \approx 6.94 \times 10^{-10}. $
By comparison, the single-QPU case is limited to approximately $ d = 9 $ and gives
$ p_L \approx 4.30 \times 10^{-8}. $
Thus, distribution improves the logical error rate by roughly a factor of 62 when the interconnect is ten times noisier than local gates. 

Figure~\ref{fig:distributed_qpu_sweep} further shows that for $ r = 1 $ and $ r = 3, $ the optimum remains at the largest tested configuration, $ q = 23 $ and $ d = 47, $ giving $ p_L \approx 2.80 \times 10^{-16} $ and $ p_L \approx 1.43 \times 10^{-13} $ respectively. For $ r = 10, $ the sweet spot shifts to $ q = 9 $ and $ d = 29. $ When $ r = 30, $ distribution is no longer beneficial, and the single-QPU configuration with $ q = 1, d = 9, $ and $ p_L \approx 4.30 \times 10^{-8}$ is optimal. These sweet spots
are conditional on the simplified assumptions $ n_{\text{phys}}(d) \approx 2d^2, N_{2Q}(d)\approx 11d^2, N_{\text{inter}} = 4d(q - 1), $ and the selected physical error rates.

\subsubsection{Distributed Bivariate-Bicycle Code Evaluation}
 The purpose of this evaluation is to compare a geometrically local topological code with a nonlocal qLDPC code under the same distributed binomial framework. In the surface-code analytical model, the primary source of inter-QPU operations are checks crossing the boundaries between code patches, so the interconnect overhead scales approximately with the boundary length. By contrast, BB codes are qLDPC codes with nonlocal check structure~\cite{TillichZemor2014QLDPC, Bravyi2024BBMemory, Berthusen2025LocalQLDPC}. Therefore, when a BB code block is partitioned across multiple QPUs, a finite fraction of the two-qubit operations in each syndrome extraction cycle may become inter-QPU operations~\cite{Chandra2026DistributedBB, Tham2025DistributedMemories}. 

\begin{figure}[t]
\centering
\includegraphics[width=0.9\columnwidth]{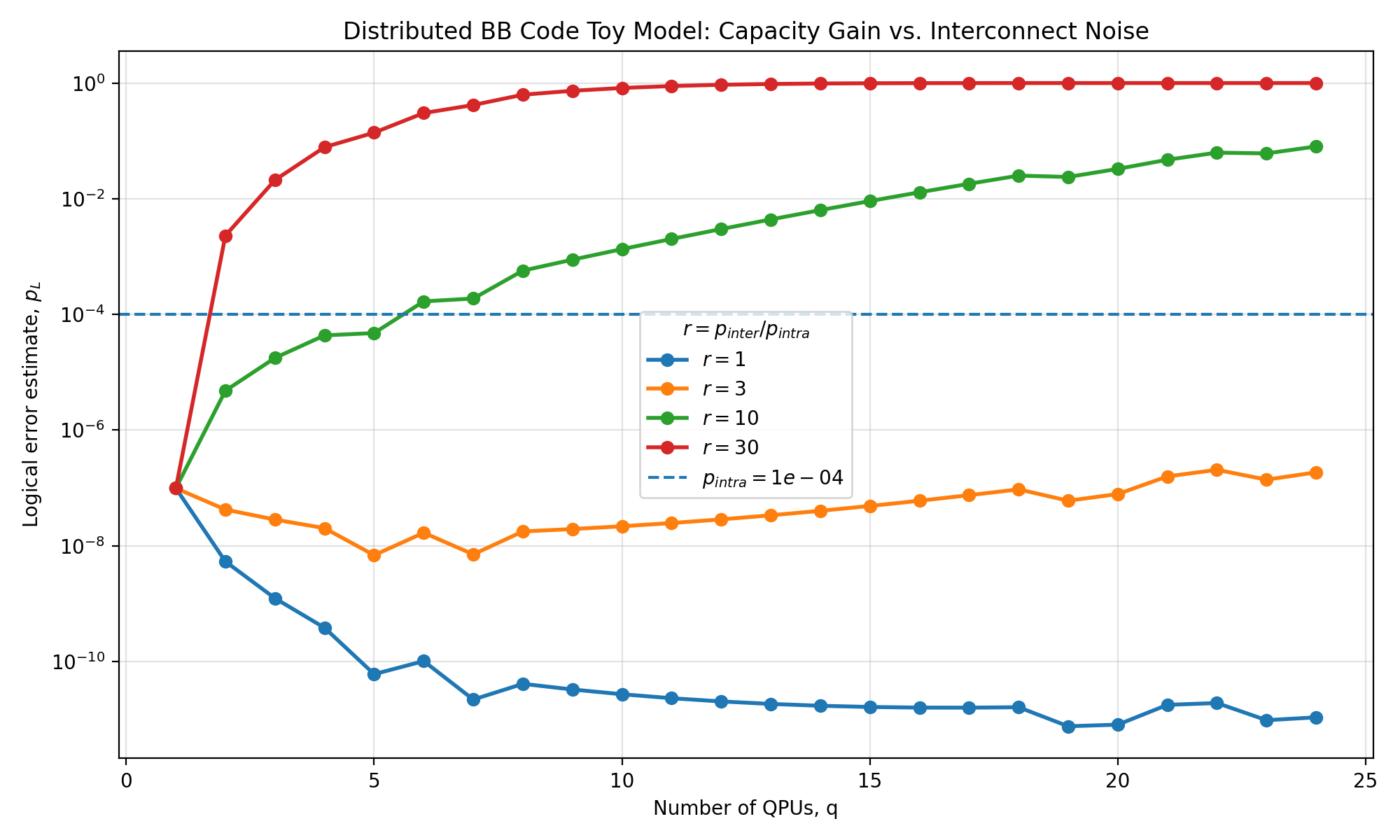}
\caption{Distributed BB Code logical error estimate: $ p_{\text{intra}} = 10^{-4}, N_{2Q}^{\text{BB}} \approx 18d^2 $ and $ n_{\text{phys}}^{\text{BB}} \approx d^2 $ with $ \lambda = 0.25. $}
\label{fig:distributed_bb_tradeoff}
\vspace{-0.1in}
\end{figure}

We use the benchmark BB instance from Table~\ref{tab:qec_codes}. The Gross BB code instance has
$ d = 12, \qquad N_{2Q} = 2592. $
These values should be interpreted as benchmark parameters rather than universal code parameters, as the two-qubit gate count depends on the syndrome extraction circuit and compilation assumptions~\cite{Swierkowska2025ECCentric, Bravyi2024BBMemory}. We approximate the two-qubit gate count using the benchmark-normalized scaling relation
$$ N_{2Q}^{\text{BB}}(d) \approx \frac{2592}{12^2} d^2 = 18d^2 $$
and also assume a constant rate physical-qubit scaling,
$$ n_{\text{phys}}^{\text{BB}}(d) \approx d^2. $$
This scaling is a proxy for the constant rate behavior of qLDPC codes~\cite{TillichZemor2014QLDPC, Bravyi2024BBMemory, yoder2025tourgrossmodularquantum}. Therefore, if each QPU supports approximately $ 200 $ physical qubits, the maximum achievable BB distance on $ q $ QPUs is modeled as
$$ d_{\max}^{\text{BB}}(q) = \left\lfloor\sqrt{200q}\right\rfloor. $$
The corresponding number of faults for logical failure is:
$$ \kappa(q) = \left\lfloor\frac{d_{\max}^{\text{BB}}(q) - 1}{2}\right\rfloor + 1. $$

The key difference from the surface code model is the scaling of $ N_{\text{inter}}. $ For a balanced partition of a nonlocal BB circuit across $ q $ QPUs, the probability that a nonlocal two-qubit interaction crosses a QPU boundary is approximately $ 1 - 1/q. $ We therefore model the number of inter-QPU operations as
$$ N_{\text{inter}}(q, d) = \lambda\left(1 - \frac{1}{q}\right)N_{2Q}^{\text{BB}}(d), $$
where $ 0 \leq \lambda \leq 1 $ is an embedding factor that represents how effectively the compiler and code layout keep checks within the same QPU. The case $ \lambda = 1 $ corresponds to an essentially random nonlocal embedding, while smaller values of $ \lambda $ represent more modular layouts where many BB checks remain intra-QPU~\cite{ShawTerhal2025MorphingBB, Chandra2026DistributedBB, Tham2025DistributedMemories}. The number of intra-QPU operations is then
$$ N_{\text{intra}}(q, d) = N_{2Q}^{\text{BB}}(d) - N_{\text{inter}}(q,d). $$
As before, the logical error probability is estimated as
$$ p_L = \mathbb{P}\left[X_{\text{intra}} + X_{\text{inter}} \geq \kappa(q) \right]. $$

Figure~\ref{fig:distributed_bb_tradeoff} shows the resulting logical error estimates as a function of QPU count for several interconnect noise ratios using $ p_{\text{intra}} = 10^{-4} $ and $ \lambda = 0.25. $ When $ r = 1, $ inter-QPU operations are as reliable as local gates, so increasing the number of QPUs improves performance by allowing a larger-distance BB code. When $ r = 3, $ there is still a distribution benefit, but the sweet spot occurs at a smaller QPU count. When $ r = 10 $ or $ r = 30, $ distribution becomes unfavorable under this BB analytical model because the nonlocal checks create an interconnect penalty that scales with a fraction of the total circuit volume~\cite{Chandra2026DistributedBB, Berthusen2025LocalQLDPC}.

The exact sweet spots in Figure~\ref{fig:distributed_bb_tradeoff} illustrate how rapidly the benefit of distribution decreases as the interconnect becomes noisier. When $ r = 1, $ the best tested configuration is $ q = 19 $ and $ d = 61, $ giving $ p_L \approx 7.57 \times 10^{-12}. $ For $ r = 3, $ the optimum moves to $ q = 5 $ and $ d = 31, $ with $ p_L \approx 6.87 \times 10^{-9}. $ For both $ r = 10 $ and $ r = 30, $ distribution is not beneficial: the best configuration is the single-QPU case with $ q = 1, d = 14, \kappa = 7, $ and $ p_L \approx 9.87 \times 10^{-8}. $ In these regimes, the added distance does not compensate for the large number of noisy inter-QPU operations.

\begin{figure}[t]
\centering
\includegraphics[width=0.9\columnwidth]{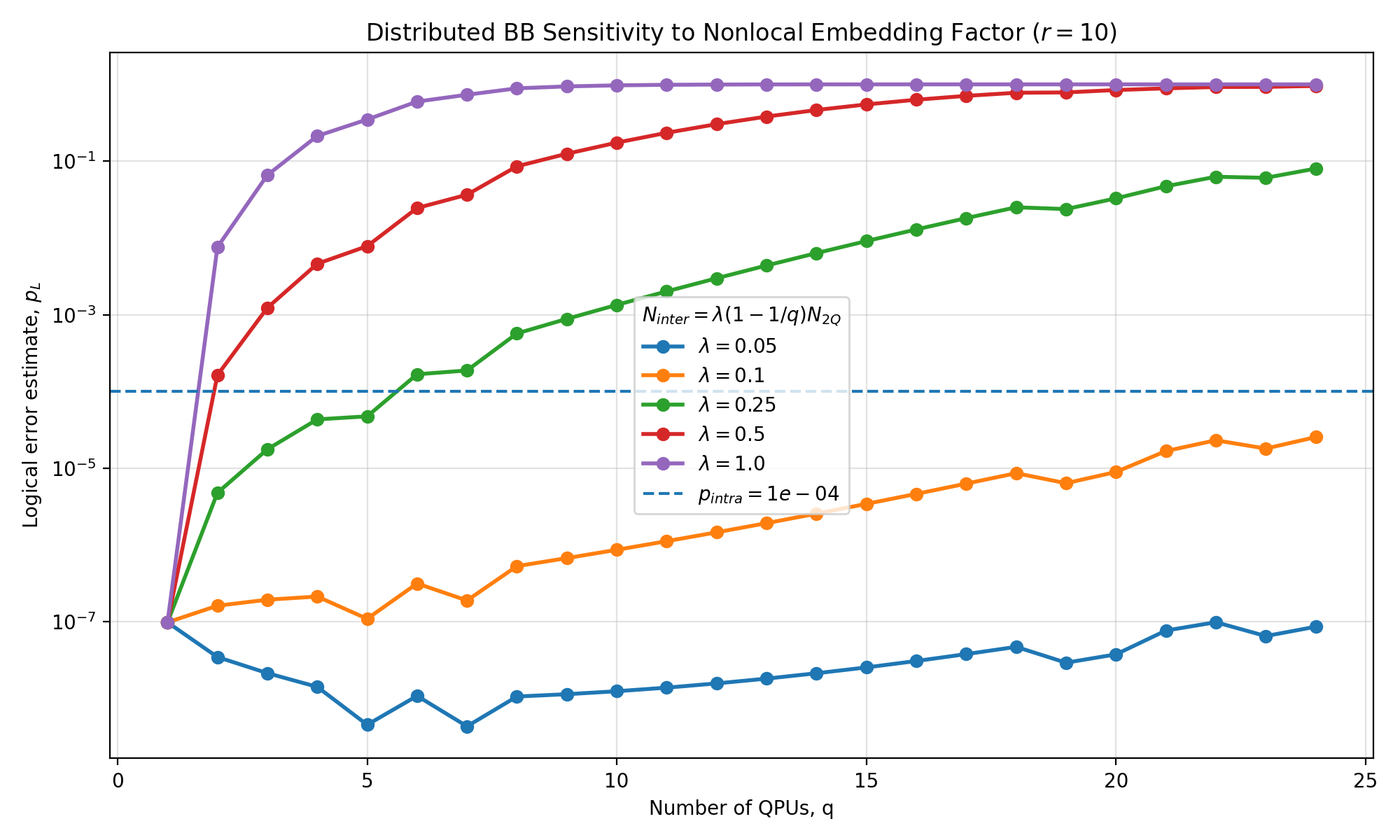}
\caption{Sensitivity of distributed BB Code: $ p_{\text{intra}} = 10^{-4} $, $ r = 10. $}
\label{fig:distributed_bb_lambda}
\vspace{-0.1in}
\end{figure}

This behavior is a sharp contrast to the surface code analytical model. For the surface code, the inter-QPU cost grows approximately as a boundary term,
$$ N_{\text{inter}}^{\text{surface}}(q, d) \sim d(q - 1), $$
whereas for the BB code it grows as a fraction of the full circuit volume,
$$ N_{\text{inter}}^{\text{BB}}(q, d) \sim \lambda\left(1 - \frac{1}{q}\right)N_{2Q}^{\text{BB}}(d). $$
Thus, the surface code can benefit from distribution when the increase in code distance dominates the boundary overhead. In contrast, BB codes require either highly reliable interconnects or a highly modular layout to benefit from distribution.

To test the importance of the embedding assumption, Figure~\ref{fig:distributed_bb_lambda} fixes $ r = 10 $ and varies $ \lambda. $ Smaller values of $ \lambda $ correspond to better partitioning, fewer cross-QPU checks, or a more modular implementation of the BB check graph~\cite{ShawTerhal2025MorphingBB, Chandra2026DistributedBB, Tham2025DistributedMemories}. The plot shows that distributed BB performance is extremely sensitive to $ \lambda. $ When $ \lambda $ is sufficiently small, distribution can still be beneficial because the larger achievable distance is not overwhelmed by interconnect faults. However, once $ \lambda $ becomes moderately large, the inter-QPU operation count grows too quickly and the logical error rate worsens under distribution.

The sweet spots in Figure~\ref{fig:distributed_bb_lambda} show that distribution becomes beneficial only when the embedding factor is very small. At $ r = 10 $ and $ \lambda = 0.05, $ the best configuration occurs at $ q = 7 $ and $ d = 37, $ with $ N_{\text{inter}} = 1056 $ and $ p_L \approx 4.37 \times 10^{-9}. $ For every tested value $ \lambda \geq 0.10, $ the single-QPU configuration with $ q = 1, d = 14, $ and $ p_L \approx 9.87 \times 10^{-8} $ is optimal.

Overall, this model suggests that distributed BB codes have a different bottleneck from distributed surface codes. For surface codes, the main question is whether the interconnect overhead along patch boundaries is small enough to justify increasing the achievable distance. For BB codes, the main question is whether the nonlocal qLDPC check graph can be embedded so that only a small fraction of checks cross QPU boundaries. Therefore, distributed BB codes are most attractive when $ p_{\text{inter}} \approx p_{\text{intra}} $ or when the effective embedding factor $ \lambda $ is very small. If inter-QPU gates are substantially noisier than local gates and $ \lambda $ is not small, then the same nonlocality that gives BB codes their constant rate properties can also make them costly to distribute~\cite{TillichZemor2014QLDPC, Bravyi2024BBMemory, Berthusen2025LocalQLDPC, Chandra2026DistributedBB}.


\section{Conclusions}
\label{sec:future}
Our analysis reveals several key insights. First, when comparing codes with similar fault thresholds, circuit volume (captured by the number of fault locations or two-qubit gates) plays a dominant role in determining logical error rates. This suggests that, in realistic hardware regimes, reducing circuit complexity can be more impactful than increasing code distance. Second, we showed that many trends observed in large-scale simulation frameworks such as ECCentric can be understood within this simplified analytical model, indicating that much of the variation across codes and hardware arises from differences in effective circuit volume~\cite{Swierkowska2025ECCentric}. Third, in distributed quantum architectures, even a small number of high-error inter-QPU operations can dominate logical failure probability, emphasizing that scalability through distribution must be balanced against interconnect reliability~\cite{PhysRevX.4.041041, Ramette2024FaultTolerant, sutcliffe2025distributedquantumerrorcorrection}. We demonstrated this tradeoff for the surface code and BB code and showed that sweet spots for QPU count and code distance can be identified given different interconnect ratios. Finally, under biased noise, codes that provide asymmetric protection against dominant error channels, such as the $ XZZX $ surface code or tailored subsystem constructions, can achieve substantial performance gains over symmetric codes, highlighting the importance of hardware-specific code design.

Future work can extend this analytical framework to incorporate more realistic error mechanisms and decoder performance. One direction is to include correlated noise models, potentially through cluster expansions or probabilistic graphical models, to better capture error propagation in structured circuits~\cite{Brown2019LeakageMitigation}. Another is to refine estimates of effective circuit volume by incorporating compilation, routing, and scheduling overheads, especially for connectivity-limited architectures~\cite{Swierkowska2025ECCentric, Wang2024Atomique}. Additionally, integrating tensor-network-based approaches could provide a systematic way to model noise propagation and decoherence in large-scale QEC circuits~\cite{ Chubb2021TensorNetworkDecoding}. Finally, combining this analytical framework with empirical calibration data from real hardware may enable rapid, hardware-specific code selection and optimization without requiring full-scale simulation~\cite{Acharya2025BelowThreshold}.

\bibliographystyle{IEEEtran}
\bibliography{refs}

\end{document}